\documentclass[iop,twocolappendix]{emulateapj}

\usepackage{amssymb}
\usepackage{amsmath}
\usepackage{color}
\usepackage{rotating}

\usepackage{lineno}
\usepackage{epstopdf}

\shorttitle{Exo-Saturn Magnetospheres}
\shortauthors{Tilley et al.}

\begin{document}

\title{Extrasolar giant magnetospheric response to steady-state stellar wind pressure at 10, 5, 1, and 0.2 AU}

\author{Matt A. Tilley, Erika M. Harnett, and Robert M. Winglee}

\affil{Department of Earth \& Space Sciences, University of Washington, Seattle, WA, USA\\Astrobiology Program, University of Washington, Seattle, WA, USA}

\begin{abstract}
    A three-dimensional, multifluid simulation of a giant planet's magnetospheric interaction with steady-state stellar wind from a Sun-like star was performed for four different orbital semi-major axes - 10, 5, 1 and 0.2 AU. We simulate the effect of the increasing, steady-state stellar wind pressure related to the planetary orbital semi-major axis on the global magnetospheric dynamics for a Saturn-like planet, including an Enceladus-like plasma torus. Mass loss processes are shown to vary with orbital distance, with the centrifugal interchange instability displayed only in the 10 AU and 5 AU cases which reach a state of mass loss equilibrium more slowly than the 1 AU or 0.2 AU cases. The compression of the magnetosphere in the 1 AU and 0.2 AU cases contributes to the quenching of the interchange process by increasing the ratio of total plasma thermal energy to corotational energy. The strength of field-aligned currents (FAC), associated with auroral radio emissions, are shown to increase in magnitude and latitudinal coverage with a corresponding shift equatorward from increased dynamic ram pressure experienced in the hotter orbits. Similar to observed hot Jovian planets, the warm exo-Saturn simulated in the current work shows enhanced ion density in the magnetosheath and magnetopause regions, as well as the plasma torus which could contribute to altered transit signals, suggesting that for planets in warmer ($>$0.1 AU) orbits, planetary magnetic field strengths and possibly exomoons - via the plasma torus - could be observable with future missions.
\end{abstract}


%\linenumbers

\section{Introduction}
\label{intro}

Planetary science has undergone a recent renaissance, from both in-situ observations of solar system missions, and the inundation of exoplanetary discoveries by recent observational efforts. Our understanding of planetary formation, evolution and general physical characteristics has until recently been based solely upon those bodies found in our local system; exoplanetary observations have informed us that our system is one in a field of near-infinite variation. Given this fact, we must work to abstractly quantify a framework for the systems to progress in planetary science. Approximately two-thirds of confirmed exoplanets are giant planets ($\ge$ 0.01 M$_{Jup}$) that orbit their host stars at relatively small distances (a $\le$ 0.5 AU) \citep{han14}. This fact serves to both challenge our notions of planetary formation and evolution, and to provide opportunities to develop new approaches to study planetary configurations to interpret and support the observational data. 

A key question addressed in recent works about planetary evolution is the rate at which mass is lost from the atmosphere. The study of aeronomy for giant exoplanets has indicated that thermal and non-thermal processes both contribute strongly to mass loss for these giant bodies \citep[e.g.][]{yelle08}.  The overall system of exoplanetary mass transport involves escaping atmospheric species from the upper planetary atmosphere and active satellites outward throughout the surrounding magnetic environment. A large amount of material escaping from the atmosphere of a planet is likely to become ionized and contribute to dynamics in the planetary magnetosphere. The inclusion of this magnetic environment is a step towards a comprehensive view of planetary systems; the magnetic environment not only contributes to mass transport, but to the potential generation of detectable radio signals that can be used for detection and characterization \citep[e.g.][]{zarka07}, as well as potential alterations of transit light curves which would allow some insight into planetary characteristics \citep[e.g.][]{vidotto11,llama13,benbal14,nichols15,alexander16}. 

Recently, the question of exomoon habitability has become quite important, as the detection of these bodies in the habitable zone of the stellar hosts is now possible \citep{helbar13}. The environment through which potentially habitable exomoons orbit can be a highly dynamic magnetized plasma system that picks up ionized species, as well as injecting energized species into the upper atmosphere of the satellites, affecting the atmospheric chemistry and mass loss rates from these smaller bodies \citep[e.g.][]{johnson10}. Alternatively, the impact of energized charged particles on the surface of an icy moon can drive the formation of organic molecules, creating the building blocks necessary for the formation of life \citep[e.g.][]{chyba00,hand07}. The balance between such processes will inform a habitable satellite versus an uninhabitable one, and therefore the inclusion of the magnetic star-planet-moon interaction is key to developing an accurate model for predicting the atmospheric state of an observed exomoon.

A magnetized stellar wind can couple directly with planetary magnetospheres through reconnection at the magnetopause and down the magnetotail \citep{dungey61} and provides external forcing through viscous interaction \citep{axford61}. The Dungey reconnection framework is known to be a significant driver of magnetospheric interaction with the solar wind, as reconnection dominates production of plasma flows. Corotation and viscous flows are thought to be as important for rapidly rotating magnetospheres with significant internal plasma sources, like Jupiter and Saturn, though it is still not clear to what extent relative to the Dungey-type magnetospheric picture \citep[e.g.][]{brice70,delamere10}. Determining the response of planetary magnetospheres to various stellar wind conditions is a primary task in comparative planetary science. Key to this task is understanding the relative contribution to the global dynamics of magnetospheric systems of both external processes, e.g. stellar wind characteristics, and internal processes, e.g. rapid corotation of heavy-ion plasma transmitted by the corotating planetary magnetic field. The imperative of understanding this balance is especially true for those planets which are subject to both significant external and strong internal processes, such as Saturn. The balance between these internal and external drivers contribute to the processes of mass loss, as well as the energization of particles throughout the magnetosphere, setting up vastly different potential orbital environments for satellites.

At $\sim$10 AU, the magnetosphere of Saturn experiences $\sim$0.01$\times$ the stellar wind dynamic ram pressure as the terrestrial magnetosphere due to decreased steady-state solar wind density, but the magnetospheric cross-section of Saturn is $\sim$400$\times$ larger. This leads to a scaling for the incident power by the solar wind on the magnetosphere of Saturn that is $\sim$4$\times$ that of Earth under similar conditions \citep[e.g.][]{zarka98}. At Earth, however, the magnetosphere is almost entirely externally driven by the solar wind due to a slower rotation combined with the lack of significant internal ion sources.

Given only the amount of forcing by the solar wind on Saturn and the scaling mentioned above, one could simply assume that Saturn's magnetospheric dynamics in response to the solar wind are much like Earth's. While the mass of Saturn is two orders of magnitude greater than Earth, the effects of gravity in Saturn's magnetosphere are similar to that of Earth's for similar scales; at one bar of atmospheric pressure at Saturn, the gravitational acceleration is just 6.5\% greater than that at Earth's surface. Saturn's magnetosphere, however, has two primary characteristics that differentiate it from Earth's: a rapid corotation with the planet itself, a factor of $\sim$2.2 greater than Earth, and a constant source of water group heavy-ions, e.g. OH$^+$ and O$_2^+$, formed from H$_2$O that is continuously injected at $\sim$4R$_S$ by Enceladus; these neutrals subsequently undergo ionization and pick-up to corotate with Saturn's magnetosphere \citep{tokar08}. At just a few percent concentration, these heavy ions can strongly affect the dynamics of the inner magnetosphere by carrying the bulk of the kinetic energy of corotation, and dominating the pressure over H$^+$ \citep[e.g.][]{thomsen10}. The injection of these heavy-ions provides a constant source of mass input to the magnetosphere of Saturn, the loss of which is drained through global, magnetospheric processes that are not currently well-understood. An open question that is relevant to the Saturn-like system discussed in this paper concerns how injected heavy ion plasma transports out of the magnetosphere with varied stellar wind external forcing.

Though internal heavy-ion effects at Saturn are key to understanding global magnetospheric behavior, the solar wind dynamic ram pressure also controls large scale dynamics, and directly influences the radio power radiated by giant planetary magnetospheres, like Saturn's Kilometric Radiation (SKR). Radio emissions can be controlled by the variations in the solar wind \citep{desch82,desch83}, which drives related large scale dynamics throughout both the dayside and nightside magnetosphere, and Dungey-like reconnection events in magnetotail or cusp regions \citep[e.g.][]{bunce05,cowley05,mitchell05}. This radio emission behavior has been observed throughout the solar system for all planets with a global magnetic field \citep{desch84}. However, there are at least two sources of power for radio emissions observed in the solar system - solar wind and rotation - the aforementioned simple scaling estimates extrapolated from observations throughout the solar system indicate that certain exoplanets may emit detectable radio emissions driven by stellar wind-magnetosphere interaction. 

Several authors have used global simulations to investigate the direct magnetic star-planet interaction (SPI) for closely orbiting hot Jovian planets which, until the Kepler mission, was the most commonly observed configuration due to selection bias of the detection methods involved (e.g. radial velocity method). For example, \citet{preusse07} used ideal MHD simulations to investigate the star-planet interactions of giant planets for varying orbital distances of 0.01-0.2 AU around a Sun-like star. The majority of the cases studied took place within the Alfv\'{e}n and sonic critical radii of the star which allows direct feedback of planetary plasma onto the star. However, the simulated case in the super-Alfv\'{e}nic region at 0.2 AU showed a planetary magnetosphere very much like the compressed dayside dipole and stretched magnetotail structure exhibited by the solar system planets with significant global magnetic fields. Ideal MHD was also used to investigate SPI for an observed, transiting exoplanet, HD 189733b \citep{cohen11a}; the models predicted reconnection events that induce a significant planetary magnetospheric mass loss on the order of 10$^{-12}$ M$_{Jupiter}$ yr$^{-1}$, and that the energy dissipated by SPI could explain modulations in the Ca II lines previously observed for the host star \citep[e.g.][]{shkolnik08}. The same model was then used to investigate the effects of a coronal mass ejection (CME) on a hypothetical hot gas giant, $\sim$1.5R$_{Jup}$, which yielded magnetospheric predictions involving significant changes to planetary magnetospheric response. Field-aligned current systems were altered from the two lobe model, common to Earth and the giant planets under nominal conditions in our solar system, to resemble Alfv\'{e}n-wave driven wings around the planet, similar to those formed at Io and Europa, which has implications on the processes of energy distribution from the CME into the magnetosphere \citep{cohen11b}.  \citet{llama13} used a 3D MHD stellar wind model and an analytical planetary bow shock model to investigate the potential for a dense planetary bow shock region to contribute to transit observations for HD 189733 b, and found that indeed for a star with similar composition to the Sun, such absorption of the emitted stellar \ion{Mg}{2} is possible given the authors' assumption of low temperature magnetosheath plasma. \citet{benbal14} used a particle-in-cell (PIC) code to investigate the effects of a potential satellite produced plasma torus on transit curves for WASP-12 b and HD 189733 to explain the early ingress identified in observations \citep{fossati09,haswell10,nichols15,alexander16}. These findings open an exciting new vista on planetary science. The findings suggest that it is possible to extract exoplanetary characteristics from transit light curves, with appropriate modeling and data analysis. In combination with radio emissions, we could eventually piece together a more complete understanding of the exoplanets that have been confirmed as instrumentation and modeling intersect and improve.

With the exception of \citet{benbal14}, the efforts mentioned above were carried out using single fluid, 3D ideal MHD models, and developed an initial global analysis of stellar wind influence on the planetary magnetosphere of hot gas giants. These simulations differ from the ones used in the present work in their use of ideal MHD simulations and focus on Jovian planets in very hot orbits ($\sim$0.02 AU). Ideal MHD simulations, however, fail to capture some important physics driven by the mass differences between heavy and light ions but are captured by the multifluid model \citep[e.g.][]{winglee09}. This paper presents the results for 3D multifluid simulations of Saturn-like planets, which are relatively smaller and magnetically weaker than the hot Jovians mentioned above. The investigation includes the Saturn-like planet at various semi-major axes, the closest of which is $\sim$0.2 AU, which is approximately one order of magnitude farther from the host star than the previously mentioned simulations. A Saturn-like planet was chosen as baseline instead of a Jupiter-like for three main reasons. First, the recent wealth of Cassini data provides a more comprehensive baseline for the validation case at 10 AU so that the starting point is data-driven. Second, the weaker planetary magnetic field allows for a less computationally intensive load for the 3D simulation - the Alfv\'{e}n speed calculated at the inner boundary often sets the timestep of the simulation. Lastly, a smaller injection rates produce a more stable satellite plasma torus with less mass lost to the inner boundary due to pressure gradients. 

The work herein is a next step for the study of stellar-extrasolar gas giant magnetospheric interactions, investigating potential plasma populations including ionospheric outflows of heavy-ions or injection by natural satellites. The heavy ions can dominate the corotational kinetic energy of the magnetosphere at a rate of only a few percent of the total plasma population. High heavy-ion injection rates also have an influence on magnetospheric pressure balance with the stellar wind by increasing the internal plasma pressure of the magnetosphere \citep[e.g.][]{pilkington15}. Building on the previous works, the efforts in this paper will contribute to a ground truth for warm gas giant magnetospheric response to stellar wind. We use our simulations with this regard. 

The 3D global multi-fluid, multi-scale model, is outlined in Section 2, along with our boundary and initial conditions. Section 3 contains the discussion and results of our simulations. A summary of our findings and conclusion is offered in Section 4.

\section{Model Details}
\label{model}

\subsection{3D Multifluid Equations}
\label{multifluid}

The 3D multifluid model separately tracks multiple, individual ion species, denoted below by subscript $\alpha$ in Eqs. (1) - (3), the conservative forms for mass, momentum and pressure (thermal energy density). The three fluid ion species have fixed mass-to-charge ratio of 1, 18, and 32 amu C$^{-1}$ as the multifluid code cannot differentiate the physics of multiply-ionized species from singly-ionized species with identical m/q ratios. The assumption is that the total population is dominated by singly ionized species, which is reasonable for ions present at distances in a warm orbit near the sun \citep[e.g.][]{zurbuchen08}.

\begin{gather}
	\frac{\partial \rho_{\alpha}}{\partial t} + {\bf \nabla} \cdot ( \rho_{\alpha} {\bf v}_{\alpha} ) = 0\label{masscons}\\
	\rho_{\alpha} \frac{d {\bf v}_{\alpha}}{d t} = q_{\alpha} n_{\alpha} ( {\bf E+ v_{\alpha} \times B}) - {\bf \nabla} P_{\alpha} - \left( \frac{GM_P}{R^2}  \right)\rho_{\alpha}{\bf \hat{r}}\label{momcons}\\
	\frac{\partial P_{\alpha}}{\partial t} = -\gamma {\bf \nabla} \cdot (P_{\alpha} {\bf v}_{\alpha}) + (\gamma - 1) {\bf v}_{\alpha}\cdot {\bf \nabla} P_{\alpha}\label{presscons}
\end{gather}

\noindent where $\rho_{\alpha}$ is the mass density, ${\bf v}_{\alpha}$ is the bulk velocity, $n_{\alpha}$ the number density and $q_{\alpha}$ the charge. $G$ is the gravitational constant, $M_P$ and $R$ are the planetary mass and radial distance from the planet (Saturn, for the present work), ${\bf E}$ is the electric field, and ${\bf B}$ is the magnetic field. $P_{\alpha}$ is the pressure associated with each ion species, $\alpha$, and $\gamma$ is the adiabatic index (5/3). 

Electrons in the model are treated as a charge-neutralizing, mass-conserved fluid, and assumed to be in steady-state or drift motion (i.e. $dv_e/dt=0$), which simplifies Eq. \ref{momcons} for electrons to

\begin{gather}
	{\bf E} + {\bf v}_e \times {\bf B} + \frac{\nabla P_e}{e n_e} = 0\label{elecmomcons}.
\end{gather}

\noindent Given quasi-neutrality, and the definitions for current density, ${\bf J}$, and Amp\'{e}re's Law, the pressure completes the description of electron dynamics:

\begin{gather}
	    n_e = \sum\limits_{i} n_i,\hspace{2em}{\bf v}_e = \sum\limits_{i}\frac{n_i}{n_e} {\bf v}_i - \frac{{\bf J}}{e n_e},\hspace{2em} {\bf J} = \frac{1}{\mu_0}{\bf \nabla \times B}\label{elecclose}\\
	\frac{\partial P_{e}}{\partial t} = -\gamma \nabla \cdot (P_{e}{\bf v}_{de}) + (\gamma - 1){\bf v}_{e}\cdot \nabla P_{e}\label{elecpress}
\end{gather}

\noindent where $e$ is the magnitude of electron charge, $n_e$ is the electron number density, and $P_e$ is the pressure of the electron fluid. One can then substitute the Eqs. \ref{elecclose} into Eq. \ref{elecmomcons} to obtain the modified Ohm's law,

\begin{gather}
	{\bf E} = -\sum_i \frac{n_i}{n_e} {\bf v}_i {\bf \times B} + \frac{{\bf J\times B}}{e n_e} - \frac{1}{e n_e} {\bf \nabla} P_e + \eta {\bf J}\label{ohms},
\end{gather}

\noindent where $\eta$ is the resistivity, which is added only in the ionosphere to allow finite conductivity. Everywhere else, $\eta$ is zero so there is no anomalous resistivity in the model. The strength gained by modeling the electron and ion-species separately is that the model retains Hall and pressure gradient terms in the modified Ohm's Law, Eq. \ref{ohms}, which are sufficient to drive reconnection. One can substitute Eq. \ref{ohms} into Eq. \ref{momcons} to obtain the ion momentum, 

\begin{gather}
	\rho_{\alpha} \frac{d {\bf v}_{\alpha}}{d t} = q_{\alpha} n_{\alpha} \left( {\bf v}_{\alpha} - \sum_i \frac{n_i}{n_e} {\bf v}_i \right) \times {\bf B} + \frac{q_{\alpha} n_{\alpha}}{e n_e} \left({\bf J \times B} - {\bf \nabla} P_e\right)\nonumber \\
    - {\bf \nabla} P_{\alpha} + q_{\alpha} n_{\alpha} \eta {\bf J} - \left( \frac{G M_P}{R^2}  \right) \rho_{\alpha}{\bf \hat{r}}.\label{totmom}
\end{gather}

If one assumes a single ion species (single-fluid MHD) or a single velocity for all species (single-fluid, multi-species MHD), then Eq. \ref{ohms} reduces to a form inherent to ideal MHD, and the first term of Eq. \ref{totmom} disappears. It is this contribution from tracking multiple species with independent velocities that allows the multifluid model the capture ion cyclotron effects. The equations for the multifluid model were solved using a second order Runge-Kutta method on a nested grid. 

\subsection{Simulation Grid}
\label{sec:bcs}
For each simulation in the present work, the following grid parameters were kept static across all planetary simulations. The simulation's Cartesian coordinate system is such that $x$ is in the direction of a planet-star line in the equatorial plane of the planet, positive pointing away from the star. The $z$ axis is aligned with the planetary rotation axis positive towards the northern magnetic pole as the magnetic configuration of the planet is axisymmetric. The $y$ direction completes a right-handed coordinate system, and points in the direction of the tangent to planetary orbital motion. 

Five cubic, nested grids were used for each case in the study, with the innermost grid centered on the planet with an inner boundary at 2.0 $R_P$. This inner grid is $\pm$12$R_P$ in the $x$- and $y-$directions, and $\pm$6$R_P$ in the $z$-direction, with a resolution of 0.2$R_P$ in all dimensions. Each higher-order grid increases by a factor of 2, and so the outermost grid is a factor of 16 larger, but not necessarily centered on the planet. The outermost grid extends from -128$R_P$ sunward to 256$R_P$ down the magnetotail in the $x$-direction, $\pm$192$R_P$ in the $y$-direction, and $\pm$96$R_P$ in the $z$-direction. This scaling allows the multifluid model to capture dynamics across multiple scales, from a fraction of a planetary radius, up to massive structure formation down the magnetotail. 

\begin{deluxetable}{c|cccc}

\tablewidth{0.8\linewidth}

\tablecaption{Stellar wind conditions and magnetopause parameters}
\tablenum{1}

\tablehead{\colhead{} & \colhead{10 AU} & \colhead{5 AU} & \colhead{1 AU} & \colhead{0.2 AU/D/S} } 

\startdata
n$_H^+$ (cm$^{-3}$) & 0.065 & 0.26 & 6.5 & 162.5 \\
T$_H^+$ (eV) & 1.2 & 2.5 & 4.7 & 39.0 \\
P$_{dyn}$ (nPa) & 0.022 & 0.088 & 2.2 & 55.0 \\
IMF $\lvert{B}\rvert$ (nT) & 0.51 & 1.0 & 5.1 & 27.2 \\
IMF $\lvert{B}_z\rvert$ (nT) & -0.13 & -0.26 & -1.3 & -6.4 \\
IMF $\phi$ (deg) & 84.3 & 80.5 & 45.0 & 14.0 \\
R$_{MP}$ (R$_P$) & 20.9 & 16.0 & 10.1 & 5.9/6.6/3.1 \\
\enddata
\tablecomments{Stellar wind conditions and the resulting magnetopause standoff distance for each of the 4 cases discussed. The stellar wind parameters for the two additional cases at 0.2 AU (D - dense, S - slow) are the same as the 0.2 AU base case; only planetary parameters were altered, which led to differing values for the substellar magnetopause distance. $\sim$ \label{tab:swparams}}

\end{deluxetable}
\subsection{Initial Conditions}
\label{ics}

The 3D multifluid code tracks three distinct, separate ion species and a separate electron fluid. H$^+$ is present in the solar wind, and is also present in the Saturn-like planet's ionosphere and plasma torus. Two heavier species are tracked: a 18 amu ion fluid, representing potential medium mass species such as O$^+$, or H$_2$O$^+$, and a heavier fluid at 32 amu, representing more massive ions such as O$_2^+$. The planetary body for each baseline simulation case is assumed to be identical to that of Saturn, in terms of radius, mass, obliquity, rotation, magnetic field, and ionosphere so that comparisons are solely reliant upon the changing dynamic ram pressure (except for two cases at 0.2 AU discussed below). Saturn was chosen to set a strong, ground truth baseline from the wealth of Cassini observations available \citep[e.g.][]{dougherty09}. 

A Sun-like, G-type star, is assumed to generate the stellar wind for each case in this study. The stellar wind, consisting of a quasi-neutral plasma comprised of H$^+$ and electrons, is blown into the grid system from the negative $x$-direction, at a speed of 450 km/s in the direction of the positive $x$-axis, with the IMF direction calculated according to a Parker spiral. The planetary rotational axis is aligned to simulate an equinox seasonal configuration. Each case represents the simulation at a different orbital semi-major axis around the star: 10 AU, 5 AU, 1 AU, and 0.2 AU; the latter distance at 0.2 AU includes a base case, with only the effects of stellar wind pressure from the hotter orbit taken into account, and two additional studies. The additional studies include the baseline dynamic ram pressure from the stellar wind at 0.2 AU, with the addition of one case with a decreased planetary rotation rate, to investigate potential gravitational tidal effects, and one case with higher ionospheric and satellite torus plasma density to account for increased photoionization (see Section \ref{considerations}). The radial temperature dependence of the incident stellar wind is provided by the synthesis of Voyager 2 data and temperature-velocity relations, after \citet{richardson03}. Density of the stellar wind was assumed to follow a $R^{-2}$ isotropic expansion scaling relation, following Voyager 2 observations \citep[e.g.][]{belcher93}. The interplanetary magnetic field (IMF) carried by the stellar wind plasma follows the assumption of a Parker spiral, i.e. $B_r \propto R^{-2}$ and $B_{\phi} \propto R^{-1}$ - see Table 1 for details. To isolate the effect of dynamic forcing from the steady state stellar wind's pressure on the planetary magnetosphere, the orientation of the $z$-component of the IMF is held consistently in a negative orientation, or parallel to the equatorial dipole field of the Saturn-like planet. Future work is planned to address the effects of an open-type magnetospheric interaction - including Dungey-type reconnection. For each case mentioned above, Table 1 contains the relevant steady-state conditions. 

\subsection{Plasma Torus}
\label{torus}

An Enceladus-like plasma torus was included in the model, and maintained for each semi-major axis at which the simulations were run. The plasma torus was injected equally across azimuth with Gaussian cross-section, with a mean at 4 R$_P$, $\sigma$=0.5 R$_P$, which gives a volume of $\sim2\pi^2 4 R_P^3$. Ionization rates for the plasma torus were calculated for Saturn, given the expected rates for charge exchange, electron impact ionization, ionization-dissociation reactions, electron recombinations reactions, dissociative electronic recombination reactions and photolytic reactions as given by \citet{fleshman10}; these calculations are in line with Cassini observations for the H$_2O$-group species ratios reported in \citet{wilson15}. Three species were injected into the torus : H$_2$O-group ions (H$_2$O$^+$, O$^+$, OH$^+$, and H$_3$O$^+$), H$^+$, and O$_2^+$. The 18 amu H$_2$O-group ions are injected at a rate of $\sim$1.2$\times$10$^{28}$ ions/s, which correlates to the upper limits of Cassini and Hubble observations for the Enceladus plasma torus and neutral cloud source. The H$^+$ and O$_2^+$ ions are injected at a rate relative to the H$_2$O$^+$ rates of $\sim$7.3\% and $\sim$0.19\%, respectively for a total of $\sim$360 kg/s plasma injected. While these are the prominent ion species in the magnetosphere of Saturn, we assume that for the exo-Saturns, the discussion centers around elements/molecules with mass to charge ratios near 1, 18 and 32 amu C$^{-1}$. The model's simulated dynamical variation between heavy ion fluids that differ by a few percent (e.g., 16 vs. 18 amu C$^{-1}$) is small, and so these simulations can be taken to represent the behavior of general ionized species with mass to charge ratios of $\pm\sim$15\%. 

Heavier(lighter) ion species mass would contribute to higher(lower) pickup ion energies, and larger(smaller-)-radius bulk cyclotron motion for the injected ions in the model, and contribute more strongly(weakly) to the corotational energy of the system. When comparing the injected torus between the 3 masses in the present simulation, the 1 amu injected plasma is more tightly bound to the radial region at 4 R$_P$ when compared to the 18 and 32 amu ions. Corotationally-driven transport is largely influenced by the ion fluid with the greatest corotational energy - at a Saturn-like planet, the 18 amu ion group. It would be interesting to explore the inner magnetospheric configuration and mass transport reliance on both the mass of injected species and relative concentrations of those ions. Such an investigation is best performed in a future effort, as the present work is focused on the response of a Saturn-like configuration to steady-state stellar wind changes.

\subsection{Considerations at 0.2 AU}
\label{considerations}

Three additional factors were taken into account when considering the planetary configuration at 0.2 AU: 1) tidal dissipation leading to planetary rotational changes affecting the generation of the planetary magnetic field, 2) increased radiation due to stellar proximity, and 3) orbital stability of the internal source of heavy-ions, or exo-Enceladus. 

Tidal dissipation in the planet for the 0.2 AU case was considered by using a solution to determine the locking timescale, $\tau$, using the constant phase-lag method described in Appendix E of \citet{barnes13}. $\tau$ was calculated over a range tidal dissipation factors, $Q = 1\times10^5$-$1\times10^6$ where $k_2$, the Love number was set to 0.3, and initial eccentricity set to 0.3. For the lower value of $Q=1\times10^5$, the planet reached a 3:2 spin-orbit resonance, like Mercury in the solar system, after $\sim$2.08 Gyr, but never synchronously locked; for the upper bound of $Q=1\times10^6$, the planet had only doubled its spin period (from $\sim$10.7 h to $\sim$20.4 h) after 10 Gyr. In light of these results, a case at 0.2 AU was run for the scenario with a rotational period of $\sim$522.78 h, which corresponds to a 3:2 spin-orbit resonance mentioned above. Following the scaling for a planetary magnetic field from \citet{sanchez04}, $B \propto \sqrt{\omega}$, this weakens the planetary magnetic field by a factor of $\sim$7 to $\sim$3$\times$10$^{-6}$T at the equator; this case is referred to as the 0.2AU slow case.

The effect of increased stellar electromagnetic flux was considered, but for the base case at 0.2 AU, both ionospheric density at the inner boundary of 2 $R_P$ as well as the rate of ionization in the plasma torus at 4 $R_P$ were kept constant to isolate the effect of increased stellar wind dynamic ram pressure on magnetospheric dynamics. However, a separate case at 0.2 AU included an inner boundary and satellite torus density increased by a factor of $\sim$530, which is an increase proportional to the ratio of photoionization and photolysis to overall ionization for the relevant ionic species, assuming a simple isotropic R$^{-2}$ dependence on the stellar flux; this case is referred to as the 0.2AU dense case. Similarly, the conductance of the ionosphere has been increased for the latter case by the same amount, so that the ratio of conductance to mass input is maintained.

Lastly, there was some concern about the dynamical stability of an exomoon heavy-ion source over long timescales for a giant planet at 0.2 AU. Following \citet{barnes02}, one can calculate the maximum possible extant satellite mass as

\begin{gather}
	M_m \le \frac{2}{13} \left[ \frac{(f a_p^3)}{3 M_*}  \right]^{13/6} \frac{M_p^{8/3} Q_p}{3 k_{2p} T R_p^5 \sqrt{G}},
\end{gather}

\noindent where $f$ is a constant fraction of the satellite's Hill radius, $a_p$ is the planetary semi-major axis, $M_*$ is stellar mass, $M_p$ is the planetary mass, $Q_p$ and $k_{2p}$ is the tidal dissipation parameter discussed above, $T$ is the age of the system, $R_p$ is the planetary radius, and $G$ is the gravitational constant. Using Saturn's planetary parameters at 0.2 AU around a solar analog, the tidal quantities discussed above, and calculating at the age of the current solar system, the upper bound satellite mass calculated is $M_m\le$2.01$\times$10$^{23}$ kg, or 1.86$\times$10$^{3}$ Enceladus' mass and $\sim$2.25 Io's mass. Given this result, the present study is performed under the assumption that the presence of a heavy-ion source produced by an Enceladus-like satellite in the inner magnetosphere is present and stable for a giant planet in this warm orbit.

\section{Discussion and Results}
\label{results}

The model explores the effects of viscous (IMF B$_z$ consistently parallel to the equatorial planetary magnetic field), steady-state stellar wind forcing on the magnetosphere of a gas giant; the planetary magnetosphere exhibits several modified characteristics of the structure and plasma transport, including enhanced magnetotail reconnection in the compressed cases identified by the formation of thin current sheets in regions of reconnection. One obvious effect is the compression of the magnetosphere by the increase in overall ram pressure on the planetary field. Table~\ref{tab:swparams} shows the dynamic ram pressure, and the simulated magnetopause standoff distance. This compression of the magnetosphere has a strong effect on the dynamics related to mass loss from the magnetic environment such as alteration of the radial structure of the inner magnetosphere, which in turn leads to a different mass flow profile. Similarly, the compression of this magnetic boundary ensures that at some point, the boundary layer is dominated by the high densities of injected plasma from the moon at 4 R$_P$; this changes the potential optical depth of the bow shock. The details are in the subsections that follow.

\subsection{Interchange instability dampening and mass transport} \label{sec:interchange}

\begin{figure*}[!ht]
    \centering
    \includegraphics[width=0.7\linewidth]{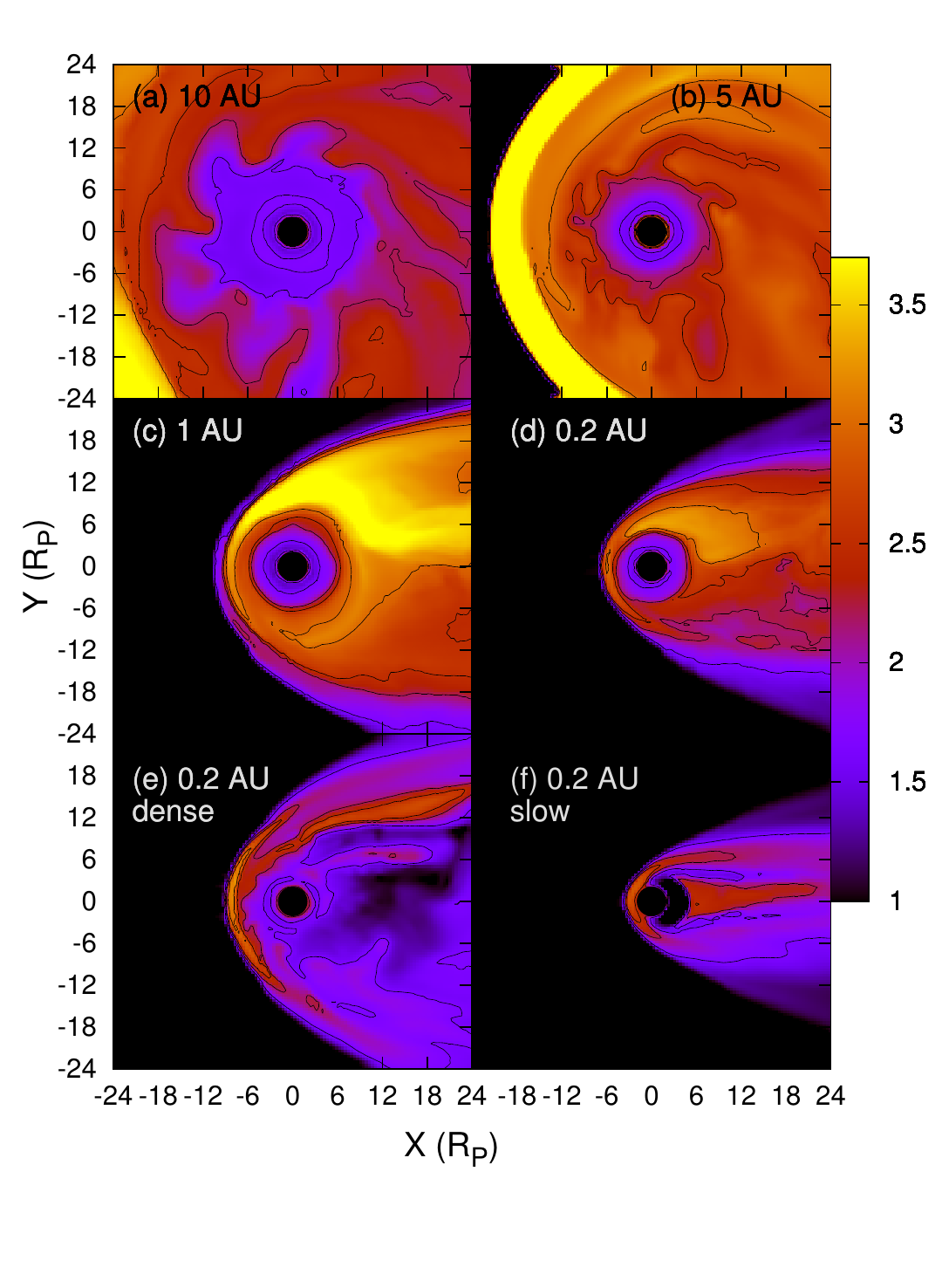}
    \caption{A view from above the planetary northern pole, showing the 18 amu ion temperature (eV) in the equatorial plane. The static temperature for the stellar wind for each case, given in Table~\ref{tab:swparams}, has been removed for improved visualization.}
	\label{fig:wtemp}
\end{figure*}

The picture for mass transport from a giant, rapidly rotating magnetosphere down the magnetotail and out into the stellar wind is not completely understood, but it is known that processes such as the centrifugal interchange instability strongly contribute to such transport for rapidly-rotating, magnetized planets \citep[e.g.][]{kivelson05,thomsen13}. The centrifugal interchange instability can be likened to the Raleigh-Taylor instability, but with the centrifugal force from planet's rapid rotation in lieu of gravity. Cold, dense heavy-ion plasma from the near-planet magnetosphere (herein, the Enceladus analog's orbit at $\sim$4 R$_P$) exchanges with the hot, diffuse plasma radially adjacent, and carries with it magnetic flux; these dense plumes possibly undergo small magnetic reconnection events and are lost from the inner magnetosphere. However, for the magnetosphere to be unstable to flux interchange, two facts must hold: the gradient of equatorial plasma flux tube content must be negative with increasing radial distance, and the thermal energy of the magnetospheric plasma must be less than that of the corotational energy \citep{hill76}. 

The marginal interchange stability criterion with respect to the flux tube content (FTC) is given by \citet{hill76} as:  

\begin{align}
    \frac{\partial}{\partial r_{eq}} \left( \frac{\rho_{eq}r_{eq}g}{B_{eq}} \right ) = 0 \label{eq:ftc_stab}
\end{align}

\noindent where $\rho_{eq}$, $r_{eq}$, and $B_{eq}$ are the equatorial mass density, radial distance and magnetic field magnitude, respectively. For a rotationally-aligned, planetary magnetic dipole, as for the present system, the field geometric factor, $g$, can be approximated as $g(L) \approx 4 L^{1/2} - 3$, where L is the equatorial radial distance in terms of planetary radii. For a configuration where the quantity in Eq.~\ref{eq:ftc_stab} is negative, the system will be unstable to interchange, i.e. if total flux tube content is decreasing with increasing radial distance from the planet.

Total flux tube content in the multifluid model was calculated by tracing closed magnetic field lines as a function of L-shell, over a colatitude range corresponding to the north and south magnetic foot points for each L-shell value, from the plasma torus at 4 R$_P$ to the identified magnetopause for each case. The contributing number of the dominant species, 18 q/m ions, for each point was added to obtain N$_{ion}$L$^2$, which is given by \citep[e.g.][]{sittler08}:

\begin{align}
	N_{ion}L^{2} = R_P^3 L^4 \int_{\theta_N}^{\theta_S} n_{ion}(L,\theta)sin^7\theta \, d\theta 
    \label{eq:ftc}
\end{align}

\noindent where $\theta_N$ and $\theta_S$ are the magnetic foot points in the north and south hemispheres per L-shell, respectively. This representation is missing a factor of 4$\pi$ when compared to the normalized form, as the present summation was manually performed at all magnetospheric longitudes and latitudes.

The thermal energy density of the magnetospheric plasma is given by the plasma pressure, $P = \sum_s n_s k_B T_s$, where $n_s$ is the number density, $k_B$ is Boltzmann's constant, and $T_s$ is the temperature summed over electrons and all ion species, $s$. The corotational energy density is given by the kinetic energy density for all ion species (electrons are neglected), $E_{cor} = \frac{1}{2}\sum_s \rho_s v_s^2$ where $v_s$ is the corotational velocity of ion species, $s$.

The panels in Fig.~\ref{fig:wtemp} show the temperature of the most abundant ion species, the 18 amu ions, to illuminate the outflowing cool, dense fingers and inflowing tenuous, hot injections for the 10 and 5 AU cases. The 18 amu species is injected by the satellite at 4R$_P$ at more than an order of magnitude higher input than any other species, and therefore will drive the inner magnetospheric dynamics - including the interchange processes. The temperature is shown from above the northern pole of the planet, for the plasma located in the equatorial plane as the centrifugal force is highest at low latitudes. It is readily apparent that the top two panels (left, 10 AU, and right, 5 AU) display the formation of the interchange fingers as have been measured in rapidly rotating magnetospheres, such as Saturn's \citep[e.g.][]{burch05,mauk05,andre05,persoon05}. The middle two panels (1 AU on left, and 0.2 AU on right) show a distinct lack of the interchange process, but like all cases of convection-dominated magnetospheres, they exhibit the well-known Vasyliunas cycle of magnetospheric plasma flow \citep{vasy83}. Interestingly, the 0.2 AU dense case (bottom panel of Fig.~\ref{fig:wtemp}) shows potential interchange-like fingers forming, as opposed to the base case in the middle right panel. To understand these behaviors, we must investigate the conditions for stability, outlined above. Overall for the 0.2 AU cases, plasma generated in the satellite torus is likely to rapidly flow out to the magnetopause,

\begin{figure}[!ht]
    \centering
    \includegraphics[width=\linewidth]{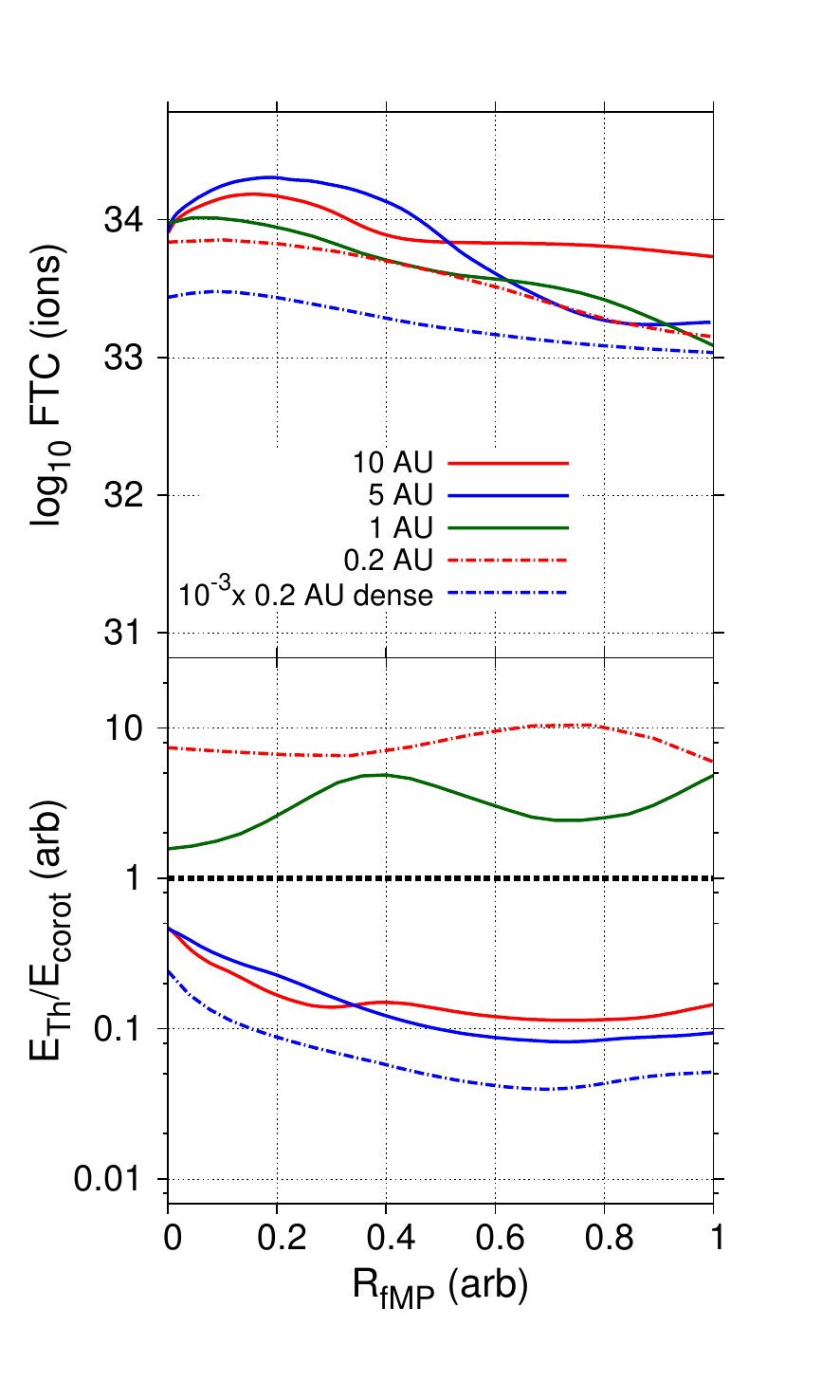}\\
    \caption{{\bf Top:} Flux-tube content (FTC) of W$^+$ ions for each case, in total ions for a givenL-shell (note the 0.2 AU dense case is scaled down by 3 orders of magnitude for comparison). {\bf Bottom:} The ratio of thermal energy density to corotational energy density. The abscissa for each plot is R$_{fMP}$, which is simply the fraction of the distance between the injected plasma torus at 4 R$_P$, and the identified magnetopause as measured in Table 1. Both top and bottom plots are averaged over 6 Saturnian rotations, and the lower plot only is an azimuthally-averaged radial profile.}
	\label{fig:interchange_conditions}
\end{figure}

\noindent and then be forced to flow down the tail, contributing directly to an increased density near the boundary of the magnetopause (see Sec.~\ref{sec:optdepth}). That mass is carried around the magnetosphere and out into the stellar wind, lost from any contribution to the inner magnetospheric processes. Likewise, for the 0.2 AU slow case, the orbit of the satellite at 4 R$_P$ takes the ion producing exomoon across the magnetopause, directly into the magnetosheath and shocked stellar wind. 

The quantities from Eq.~\ref{eq:ftc} and the ratio of total thermal to corotational energy densities are calculated for the simulation results in Fig.~\ref{fig:interchange_conditions}. Note that these quantities are time-averaged for 6 planetary rotations ($\sim$64 hours). The abscissa for this figure is given as R$_{fMP}$, which represents the fraction of the distance from the plasma torus to the magnetopause identified in Table 1, chosen to give comparable similar scaling for each case. The top panel in Fig.~\ref{fig:interchange_conditions} shows the total FTC for the 18 amu C$^{-1}$ ions in each baseline planetary configuration, and the 0.2 AU dense case. When comparing the 10 AU case in Fig.~\ref{fig:interchange_conditions} to the Cassini data baseline in \citet{sittler08}, it is noted that the present simulation overestimates the total FTC in the center of the torus and at $\sim$10 R$_P$ by approximately a factor of 2, but the peak FTC radial location ($\sim$6.5 R$_P$) and peak FTC value (1.55$\times$10$^{34}$ ions) agree well with the CAPS data. Two issues explain the differences: 1) the present simulation is injecting what is thought to be the upper end of the estimation for Enceladus' plasma input ($\sim$360 kg s$^{-1}$) from the satellite at 4 R$_P$, and 2) the simulation is run with isotropic pressure/temperature, while at Saturn there is a strong perpendicular anisotropy leading to confinement of heavy ions to the equatorial plane. Therefore there is an overestimate of ion content located on field lines at middle and high latitudes, inflating the total FTC. 

The instability criterion, Eq.~\ref{eq:ftc}, is met in all cases, though for the 10 and 5 AU cases, the condition is met in the middle magnetosphere, as opposed to the near plasma torus location for the 1 AU and 0.2 AU cases. This is consistent with Cassini observations at Saturn of a broad peak in flux tube content in the 6-8 R$_P$ range \citep[e.g.][]{sittler08,chen10}. In the 10 and 5 AU cases, we do see interchange occurring at approximately the magnetospheric radii measured by the Cassini mission (top panels of Fig.~\ref{fig:wtemp}). For the cases at 1 AU and 0.2 AU, it is apparent that the interchange-unstable condition for a negative gradient in the flux tube content is met throughout nearly the entire magnetosphere from torus to magnetopause. However, the bottom panel of Fig.~\ref{fig:interchange_conditions} shows the ratio of thermal energy to kinetic corotational energy. For a magnetospheric region to be unstable to interchange, this quantity must be less than unity. For the 10 AU, 5 AU and 0.2 AU dense cases, this second condition leaves a broad range of radial regions unstable to interchange. The cases at 1 AU and 0.2 AU baseline both show dominance by the total thermal energy, ruling out the development of the interchange fingers, as seen in Fig.~\ref{fig:wtemp}. These results imply that mass loss and flux transport in the 1 and 0.2 AU cases have been forced into a configuration akin to the terrestrial magnetosphere, i.e., driven more by the stellar wind than by corotation. 

It is likely that a Saturn-like planet with a semi-major axis between 5 and 1 AU will be unstable to interchange, but at some orbital distance that mechanism might become completely damped out. If not, before reaching 1 AU, one would expect for the magnetosphere to become stable to interchange due to the compression of the magnetopause which both increases the thermal energy of the plasma, while simultaneously decreasing the corotational energy by restricting the spatial regions of corotation. This is what happens in the present simulations at 1 and all 0.2 AU cases - the magnetosphere becomes much more controlled by the characteristics of the stellar wind as opposed to the corotational driving exhibited at larger orbital radii. 

If the interchange instability is damped out, this suggests that magnetospheric mass loss rates would be more uniform and less bursty in nature. The lack of interchange could also drive expectations for potential exoplanetary auroral observation across the electromagnetic spectrum, due to the coupling of the magnetosphere to the ionosphere of the planet \citep[e.g.][]{sittler06,nichols11,nichols12}. More discussion follows in Sec.~\ref{sec:FAC} below. 

However, the cases studied here are artificial, as the goal was to isolate the effects of the external forcing by the stellar wind pressure - the increased rates for photoionization and photolysis were not considered except for the 0.2 AU dense case. The corresponding stability criteria shown in the lower panel of Fig.~\ref{fig:interchange_conditions} suggest that the 0.2 AU dense case is dominated by the huge influx of photo-produced cold ions picked up in the satellite plasma torus. Despite the fact that the ratio of thermal to corotational energies are both linear with respect to the increased density, the corotational velocities are static with respect to radius for the 0.2 AU simulations while the azimuthally averaged temperatures are much lower for the 0.2 AU dense case due to the domination of the thermal landscape by the very dense, cold injected plasma. 

% <<<------!!!!!!!------>>> Discussion of the phase of Enceladus and magnetopause oscillations due to the increased pressure of the location. This is related to the overall pressure  balance assuming a toroidal injection similar to the assumptions made in this model.

The following question is raised: is there a location between 1 and 5 AU where the increase in photon flux leading to increased ionospheric and plasma torus densities override the effects of the compression of the magnetosphere and produce a system unstable to interchange? At 0.2 AU, the increase in photolysis and photoionization grant an overall plasma density increase by a factor of $\sim$530 when compared to Saturn at 10 AU; it is not a strict 1/R$^2$ relationship due to the fact that photoionization and photolysis are only one aspect of the ionization process, as discussed in Secs.~\ref{torus} and~\ref{considerations}. Given that, one expects at 1 AU to find an increase in total ionization of a factor of $\sim$21 over that at 10 AU. Whether such an increase is sufficient to dominate the magnetosphere of the planet at 1 AU to the extent that is seen in the 0.2 AU dense case, and at what point would such a model suggest a complete dampening of the interchange are open questions.

\subsection{Magnetospheric mass loss and mass flux}
\label{sec:flux}

\begin{figure}
    \centering
    \includegraphics[width=\linewidth]{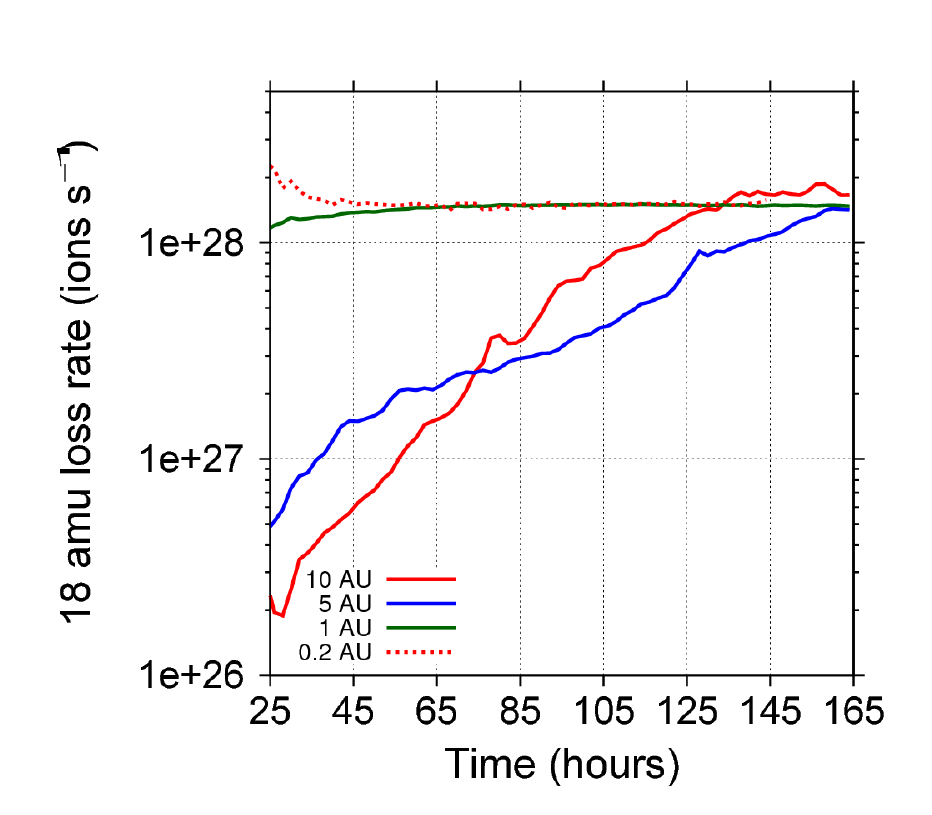}
    \caption{Plasma ion flux outflow at the outermost tailward boundary for 18 amu ions as a function of time.}
	\label{fig:flux}
\end{figure}

The process of how a body loses atmospheric atomic and molecular species, and their inherent contribution to the chemical evolution of an atmosphere, remains a question in planetary science. The question of depletion is one that is central the habitability of a planet or moon, and is governed by several thermal and non-thermal processes \citep[e.g.][]{yelle08}. Escaping species are ionized and controlled by the magnetic environment around the planet, and therefore the rates at which magnetospheric plasma is lost is related to the overall atmospheric mass loss rate for a planet. The multifluid model employed in the present work sets the density at the ionospheric boundary to be constant, replenishing any loss due to convection or pressure gradients, as such the outflow of atmospheric material is included, but not in a rigorous fashion. The primary source in the present model is that produced in the satellite torus. At equilibrium, the torus at 4 R$_P$ is injecting $\sim$360 kg/s total of 1, 18 and 32 amu C$^{-1}$ ions - $\sim$350 kg/s of which is the 18 amu C$^{-1}$ species. 

Fig.~\ref{fig:flux} is the outflow rate in ions s$^{-1}$ for the 18 amu ion species at the outermost boundaries of the simulation, outlined in Sec.~\ref{sec:bcs} for 10, 5, 1, and 0.2 AU cases. The special cases of 0.2 AU dense and 0.2 AU slow are not included in the figure as they both reach equilibrium outflow equal to their torus and ionospheric input very early in the process, similar to the 0.2 AU baseline case. It is worth noting that all baseline cases reach the same equilibrium outflow mass loss rate of $\sim$1.2$\times$10$^{28}$ ions s$^{-1}$. The presence or absence of the centrifugal interchange instability does not affect the absolute mass loss rate over sufficiently long timescales, but instead extends the relative time scale for the magnetosphere to reach equilibrium outflow during recovery after transient events. Over long timescales, all plasma injected is lost to either the inner boundary of the planet, or down the magnetotail into the flowing stellar wind. 

In the simulated magnetospheres which are unstable to interchange (10 AU and 5 AU), there is a significant relative lag time before the outflow rate of 18 amu ions matches that of the injected rate of $\sim$1.2$\times$10$^{28}$ ions s$^{-1}$. After the initial simulation equilibrated from the passage of the stellar wind at approximately 25 hours, the 5 AU case shows a lower rate of change for the ion outflow compared to the 10 AU case; the former takes approximately 140 hours to reach the equilibrated outflow rate, whereas the latter does so in approximately 100 hours, despite a factor of $\sim$2.5 lower starting outflow rate. Therefore, not only does the presence of interchange affect time scale to reach outflow equilibrium, but the results suggest that the strength of interchange may affect the features of the mass loss rate as the magnetosphere 'refills' and approaches equilibrium.

This agrees with the conditions shown in Fig.~\ref{fig:interchange_conditions}, which indicates that while both the 5 AU and 10 AU cases are unstable to interchange, the window of instability for the 5 AU case is narrower than that of 10 AU, as seen by the distance between the peak and identified magnetopause for the 18 amu species flux-tube content (FTC) in the upper panel. However, the gradient for the 5 AU case is larger, overall. The 10 AU case shows a window at distances greater than $\sim$0.18 R$_{fMP}$. Both cases are considered unstable to interchange throughout the magnetosphere, by the thermal to corotational energy ratios. This narrower window for the 5 AU case suggests a magnetosphere less unstable to interchange despite the steeper gradient, and therefore a less significant rate of increase in the rate of ion outflow occurs as seen in Fig.~\ref{fig:flux}; the simulation suggests the 5 AU magnetosphere requires a longer time to reach the equilibrium outflow rate than the 10 AU magnetosphere. The cases for 1 AU and 0.2 AU are both stable against the interchange process, and reach the equilibrium outflow rate rapidly after initial passage of the stellar wind.

The outflow rates in Fig.~\ref{fig:flux} suggest that stronger external forcing will drive a higher outflow rate after initial passage of the stellar wind. That could be likened to the passage of a coronal mass ejection (CME) or corotating interaction region (CIR) in an otherwise quiescent period; this suggests that planets with magnetospheric configuration that have sufficient radial extent to be unstable to interchange will respond initially with lower outflow. The rate of increase for the outflowing mass flux is related to the strength of the interchange instability for that magnetospheric configuration. For planetary orbital distances with increasing stellar wind dynamic ram pressure, the rate of mass loss acceleration will vary according to the overall size of the magnetosphere, i.e. the pressure balance between the stellar wind and the corotating, plasma-laden planetary magnetic field. 

For decreasing semi-major axes, the increase in stellar wind dynamic ram pressure eventually compresses the magnetosphere to the point where it is no longer unstable to interchange, bringing the system from one with mass loss dominated by internal processes to one driven by external processes. This compression forces plasma species out of the system in a more rapid fashion as the corotating region of the magnetosphere is more tightly bounded by the flow of the stellar wind and corotating plasma is lost due to the viscous interaction. The magnetospheres that are dominated by internal processes are more robust against rapidly changing stellar wind conditions, which would contribute to a more stable rate of overall mass loss from the system.

\begin{figure*}[!ht]
    \centering
    \includegraphics[width=0.75\linewidth]{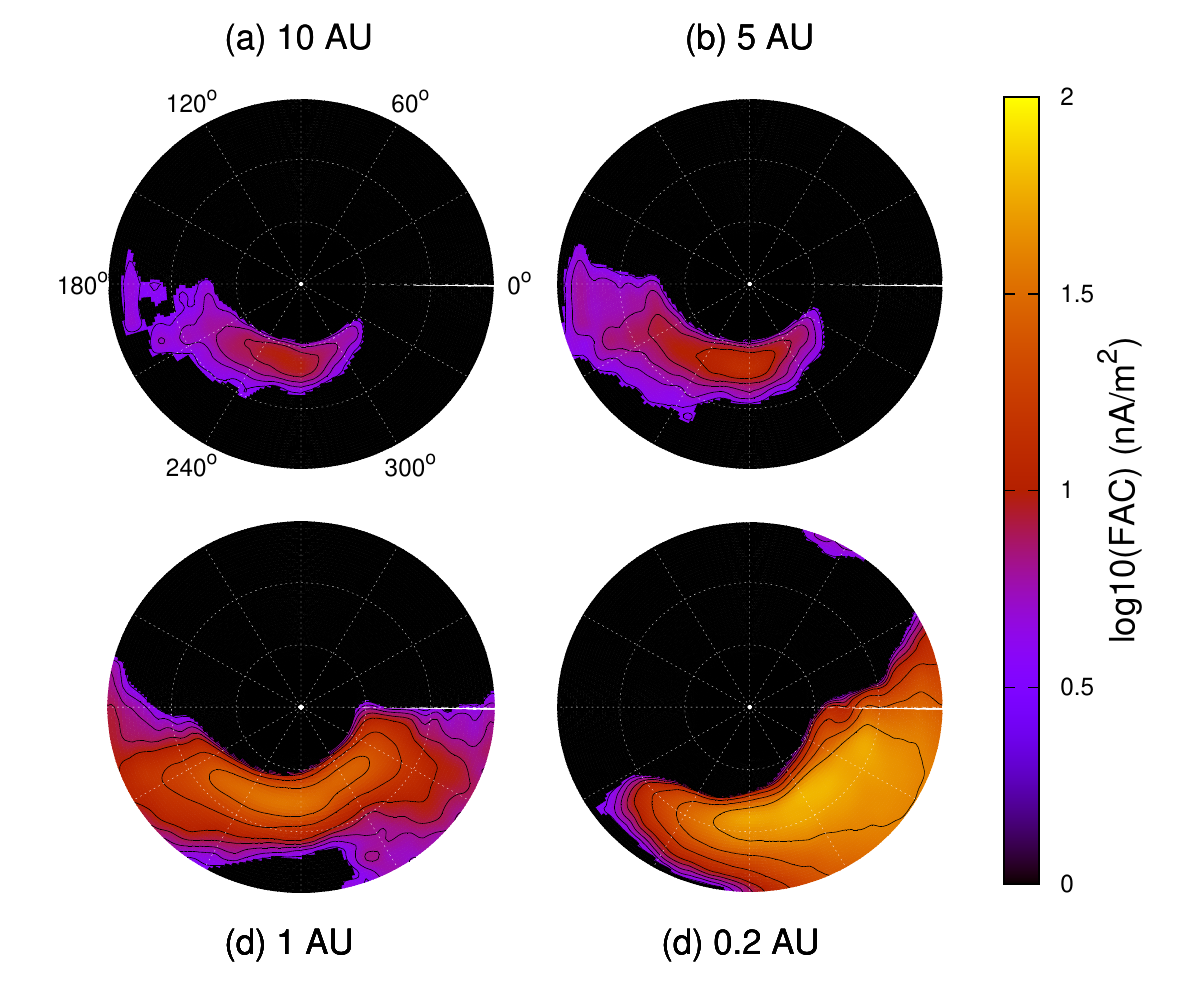}
    \caption{Time-averaged projections of upward-flowing field-aligned currents (FAC) at an altitude of 2.2 R$_P$ in the northern hemisphere of the Saturn-like planet. The radial distances correspond to 75(15), 60(30), and 45(45) degrees (co-)latitude. 0$^{\text{o}}$(180$^{\text{o}}$) points to local noon(midnight). The contours correspond to log$_{10}$ FAC values of 0.6, 0.7, 0.8, 0.9, 1.0, 1.2, 1.4, and 1.6 nA m$^{-2}$.}
	\label{fig:polarFAC}
\end{figure*}

\subsection{Field Aligned Currents and Auroral Radio Emissions}
\label{sec:FAC}

The multifluid model does not simulate kinetic processes (e.g., electron cyclotron instabilities) inherent in generating auroral radio emissions (ARE), but it does capture large-scale current systems and has validated heritage accurately simulating planetary auroral activity \citep[e.g.][]{harnett10}. In general, the power of planetary radio emission for planets with global magnetic fields has been suggested to respond directly to increased forcing from the stellar wind for rapidly rotating, giant planets like Saturn \citep[e.g.][]{desch82,desch83}. \citet{kimura13} found that the peak flux density of Saturn's ARE exhibited a positive correlation with the dynamic pressure of the solar wind on the timescale of the solar cycle. The behavior of increasing power output of planetary radio emissions seems to directly correlate to the incident kinetic and magnetic pressures on the magnetosphere for all magnetized planets in the solar system with observed non-thermal, magnetospheric radio emissions \citep[e.g.][]{desch84}. It is reasonable to expect other magnetized bodies in all planetary systems with significant stellar wind to follow a similar behavior \citep{zarka07}. 

While transient solar wind events (e.g. reconnection at the magnetopause and resulting magnetotail dynamics) can drive increased behavior in radio emission, we focus on forcing of the magnetosphere by increasing the steady-state dynamic pressure of the stellar wind. The multifluid model predicts a clear increase in FAC magnitude with increased stellar wind dynamic ram pressure, which is in accordance with the increased radio flux density observed during the solar cycle at Saturn \citep{kimura13}. However, this is counter to what one expects through a scaling study of corotation enforcement for giant magnetospheres \citep[e.g.][]{nichols11,nichols12}. In general, one expects less torque required to maintain the angular momentum of corotation for a smaller (more highly compressed) magnetosphere, and therefore a lower magnitude current system is expected.

In the present study, we explicitly hold the IMF B$_Z$ component to be negative (parallel to the equatorial planetary dipole field), and therefore Dungey-type, equatorial magnetotail reconnection is not observed in the simulations. However, plasma is lost from the inner magnetosphere down the tail for all simulated cases, with strong interchange outflow expressed for the 10 and 5 AU cases, and a continuous loss of plasma 'bubbles' for all cases \citep{kivelson05}, that leaves depleted flux tubes flowing planetward as they return on the dayside from down-tail. It can be seen in Fig.~\ref{fig:wtemp} for the 1 AU and 0.2 AU cases that plasma on the dayside is heated significantly through the return flow process, and becomes supercorotational. 

We propose that the cases simulated in the present work at 1 AU and 0.2 AU have been compressed across a threshold from being a corotationally driven magnetosphere - like that at Jupiter and Saturn - to one dominated by the increased forcing from the stellar wind. This increased forcing has led to higher tension in the magnetotail, and therefore a stronger response in return flow from down the magnetotail, which flows back towards the planet in a constant state of supercorotation. This supercorotating return flow (which is consistent with the Vasyliunas cycle return flow) was predicted and observed at Saturn, and has been proposed to strongly influence planetary auroral emission \citep[e.g.][]{talboys09,masters11}. 

Fig.~\ref{fig:polarFAC} shows a polar projection of the simulated upward-flowing, time-averaged field aligned currents above the northern hemisphere of the planet at an altitude of $\sim$2.2 R$_P$ for each of the four base cases. Note that the location and magnitude of the baseline case at 10 AU is in good agreement with measurements made by Cassini when projected along planetary dipole magnetic field lines to the visible auroral region of Saturn @ $\sim$1.02R$_P$ \citep[e.g.][]{talboys11}. The peak field-aligned current (FAC) at 10 AU is $\sim$7.2 nA m$^{-2}$, with a total magnitude of $\sim$0.82 MA rad$^{-1}$ at a colatitude of $\sim$19.1$^\text{o}$ at 2.2 R$_P$. Predicted peak currents for the warmer orbits are 1.54, 3.82, and 9.90 MA rad$^{-1}$ for the 5, 1, and 0.2 AU cases, respectively. The 0.2 AU dense case was similar to the 0.2 AU baseline in terms of latitude, though exhibited lower magnitude. The 0.2 AU slow case did not exhibit any discernible field-aligned currents in the model - as the weakened magnetic field was compressed completely to the inner boundary of the simulation, and injected plasma from the satellite was lost directly to the stellar wind for more than half of the torus volume. 

Following \citet{zarka07}, we can calculate the anticipated median radio power output based solely on the incident kinetic power on the Saturn-like planet at 0.2 AU - $\sim$7.2$\times$10$^{10}$W - a value that is nearly twice that of Jupiter's decametric median radio power. That being said, the incident IMF Parker spiral geometry at 0.2 AU for a Sun-like star as in the present work is expected to produce a lower incident Poynting flux onto an already compressed magnetosphere. \citet{zarka07} gives the dissipated power as:

\begin{gather}
    P_d = \epsilon K \left( V B_{\perp}^2/\mu_0 \right) \pi R_{MP}^2, \label{eq:pdiss}
\end{gather}

\noindent where $\epsilon$ is a reconnection efficiency of 0.1-0.2, $K$ is a function that is related to the reconnection in response to the magnetospheric state - open or closed, $V$ is the incident stellar wind speed, B$_{\perp}$ is the measure of the IMF perpendicular to the direction of stellar wind flow, $\mu_0$ is the vacuum permeability, and R$_{MP}$ is the magnetopause standoff distance. For a Saturn-like dipole configuration, $K$ is given by $\cos^4 \left(\theta / 2\right)$, where $\theta$ is the angle between the IMF embedded field and the planetary dipole. If we calculate the value of Eq.~\ref{eq:pdiss} for our Saturn-like planet at 0.2 AU (with the IMF values from Table 1), and compare with the value for Jupiter's decametric median radio power reported in \citet{zarka07}, we obtain a ratio of $\sim$5.5$\times$10$^{-3}$ which gives us a predicted power of $\sim$1.1$\sim$10$^8$W. This value is more in line with Saturn's kilometric radiation (SKR) output. It is still not clear which of the two incident powers - kinetic or magnetic - is the primary driver of emitted radio power for magnetized planets, though relative efficiencies have been suggested. 

Another feature in Fig.~\ref{fig:polarFAC} is the latitudinal position of the field-aligned currents for each case. Field-aligned currents are part of a complete circuit, with currents running in a loop from the ionosphere of the planet, along a field line down to the equatorial plasma sheet, radially to or from the planet along the plasma sheet, and then back along a field line to close in the ionosphere. It is these radial, equatorial currents which enforce corotation of the magnetosphere, due to the J$\times$B force, as seen in Eq.\ref{totmom}. The location of these field aligned currents falls within areas of subcorotation in the magnetospheres, but this is unlikely to be the cause for the magnitude increase observed in the simulation output. An extensive scaling study was performed in \citet{nichols11,nichols12} regarding this current system for Jovian planets, and its results suggest that for a purely corotational magnetosphere, compression leads to lowered corotation-enforcing field aligned current systems. In Fig.~\ref{fig:polarFAC}, the opposite is seen - with higher compression leading to an increased FAC magnitude. This suggests that the current system seen in the present simulations is not solely produced as a part of the corotational system. 

With increasing stellar wind ram pressure, the magnetopause is compressed which disallows the corotation of inner magnetospheric plasma at distances beyond this boundary, but leads to a high shear flow at this boundary between open and closed field lines. For the 10 AU case, the peak of the FAC occurs at $\sim$70.9$^\text{o}$ latitude at 2.2 R$_P$, which corresponds to an equatorial distance of $\sim$20.5 R$_P$ which is in the middle to outer dayside magnetosphere, indicating a likely corotational source. In Fig.~\ref{fig:polarFAC}, the FAC peak for the latitude for the 1 and 0.2 AU cases is located at $\sim$65.8$^\text{o}$ and $\sim$55.9$^\text{o}$, respectively, which correspond to the corotational regions at $\sim$13.1 and $\sim$7.0 R$_P$- approximately coincident with the dayside flank magnetopause location in each case. There is increased upward current at lower latitudes approaching midnight, and the high-latitude inner boundary of the upward FAC plotted for the two cases at 1 and 0.2 AU agrees well with the separatrix between the open and closed field lines. Taking this last point into account, along with the star-ward supercorotation seen in the Vasyliunas cycle return flow, and the increased stellar wind ram pressure at these orbital distances, the increase in the upward flowing FAC is likely generated by the shear flow near the magnetopause boundary.

It should also be noted that latitudinal extent of the current distributions plotted in Fig.~\ref{fig:polarFAC} correspond to a limitation of the simulation - namely that the simulation was run with lower resolution ($\sim$0.2 R$_P$) in the inner magnetosphere and at the magnetopause boundaries. This limitation washes out the finer structure that one expects for these systems. 

\begin{figure*}[!ht]
    \centering
    \includegraphics[width=0.75\linewidth]{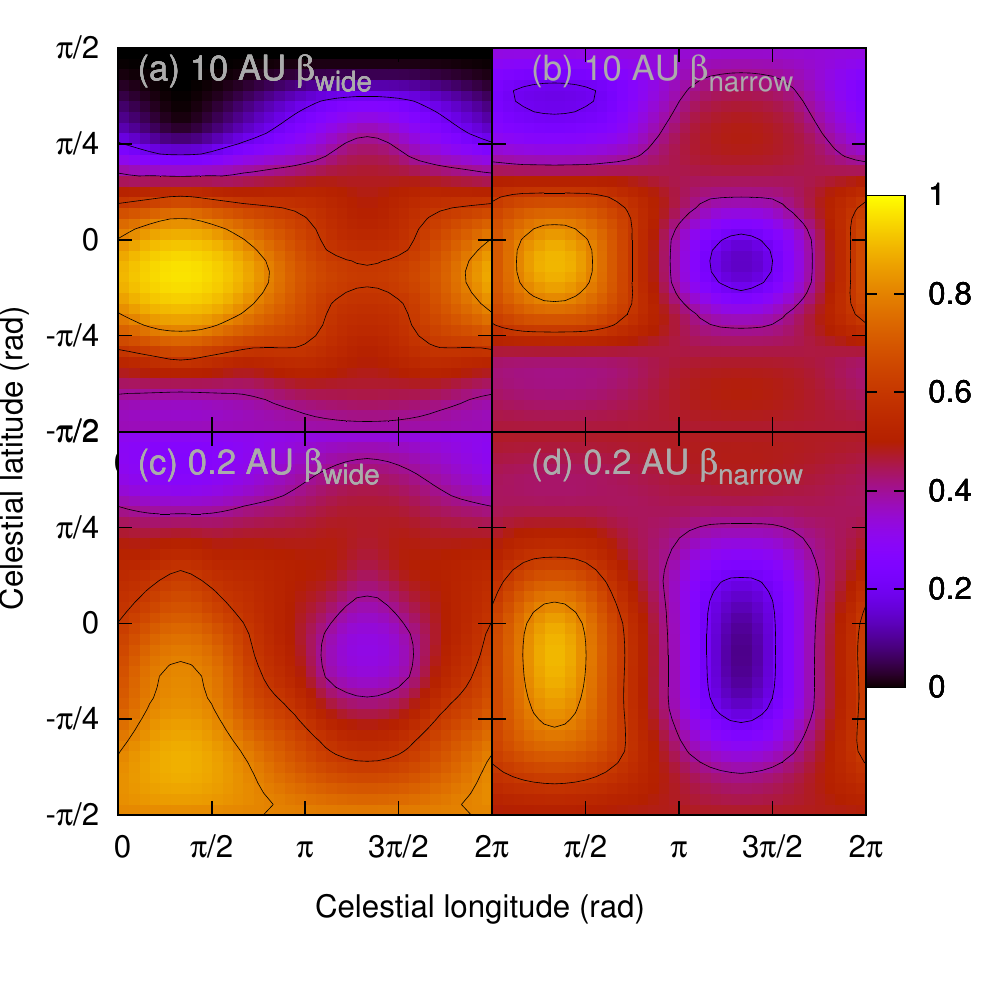}
    \caption{Heat map generated by radio emission projected onto a celestial sphere centered on the planet. {\bf Left column:} emissions with a wide beaming angle (90$^o$/60$^o$ for the northern/southern hemispheres) for the planetary configurations at (a) 10 AU, and (c) 0.2 AU. {\bf Right column:} narrower beaming angle emissions (65$^o$/45$^o$ for the northern/southern hemispheres) for the planetary configurations at (b) 10 AU, and (d)0.2 AU.}
	\label{fig:exosat_radio}
\end{figure*}

The predicted latitudes of the field aligned current systems in Fig.~\ref{fig:polarFAC} allow for an abstract spatial understanding of how such planetary configurations would project their radio signals into space. Fig.~\ref{fig:exosat_radio} shows a heat map for the emitted planetary radio signal onto a celestial sphere around the planet for the 10 AU and 0.2 AU cases (see Appendix~\ref{app:exosat}). The axes defining latitude and longitude for the planets are aligned with the magnetic moment, and for the Saturn-like planet, the rotational axis as the field is axisymmetric. The emissions were generated between 04:00 and 16:00 planetary local time (PLT), corresponding to Cassini observations of Saturnian Kilometric Radiation (SKR) emission at Saturn \citep{cecconi09}, and were latitudinally symmetric in the northern and southern hemispheres. The top row corresponds to the planetary configuration at 10 AU, and the bottom at 0.2 AU. The left column corresponds to a relatively wider beaming angle for the emitted radio power for each of the northern and southern hemispheres (see Appendix~\ref{app:exosat}), and the right column corresponds to a relatively narrow beaming angle. The two selected beaming angles correspond to 400 kHz and 200 kHz (left and right columns, respectively in Fig.~\ref{fig:exosat_radio}) and as measured by \citet{cecconi09} as an illustrative example of the projected geometric differences. Note that the hemispherical asymmetry in Fig.~\ref{fig:exosat_radio} is due to the different beaming angles measured in \citet{cecconi09}, rather than emission location. Symmetric emission altitudes and latitudes were assumed.

For the wide angle emissions in the left column, it is noteworthy that the projected radio emission is quite dissimilar for the 10 AU and 0.2 AU cases, the difference being due to the latitudinal variation between the FACs generating the radio emissions. If the combination of rotational (in this case of an axisymmetric magnetic dipole) and orbital inclination of the planet was aligned so that line of sight to our point of observation (e.g. Earth) lay in the region of high emitted power overlap - 'edge on' for the 10 AU case and 'face on' or poleward for the 0.2 AU case - radio signals with power on par with Jupiter decametric radiation (or higher) could be detected. The signals for the 10 AU cases - both wide and narrow beaming angles - are similar to a lighthouse beam, and would be expected with periodicity matching that of the orbital period for the planet. However, for the 0.2 AU wide beaming angle case (lower left panel of Fig.~\ref{fig:exosat_radio}), the situation is more interesting. Note that for a Saturn-like planet with an oppositely oriented magnetic dipole, the locations of upward- and downward-flowing current systems would be reversed, ultimately changing the projection onto the celestial sphere.

\begin{figure*}[!ht]
    \centering
    \includegraphics[width=\linewidth]{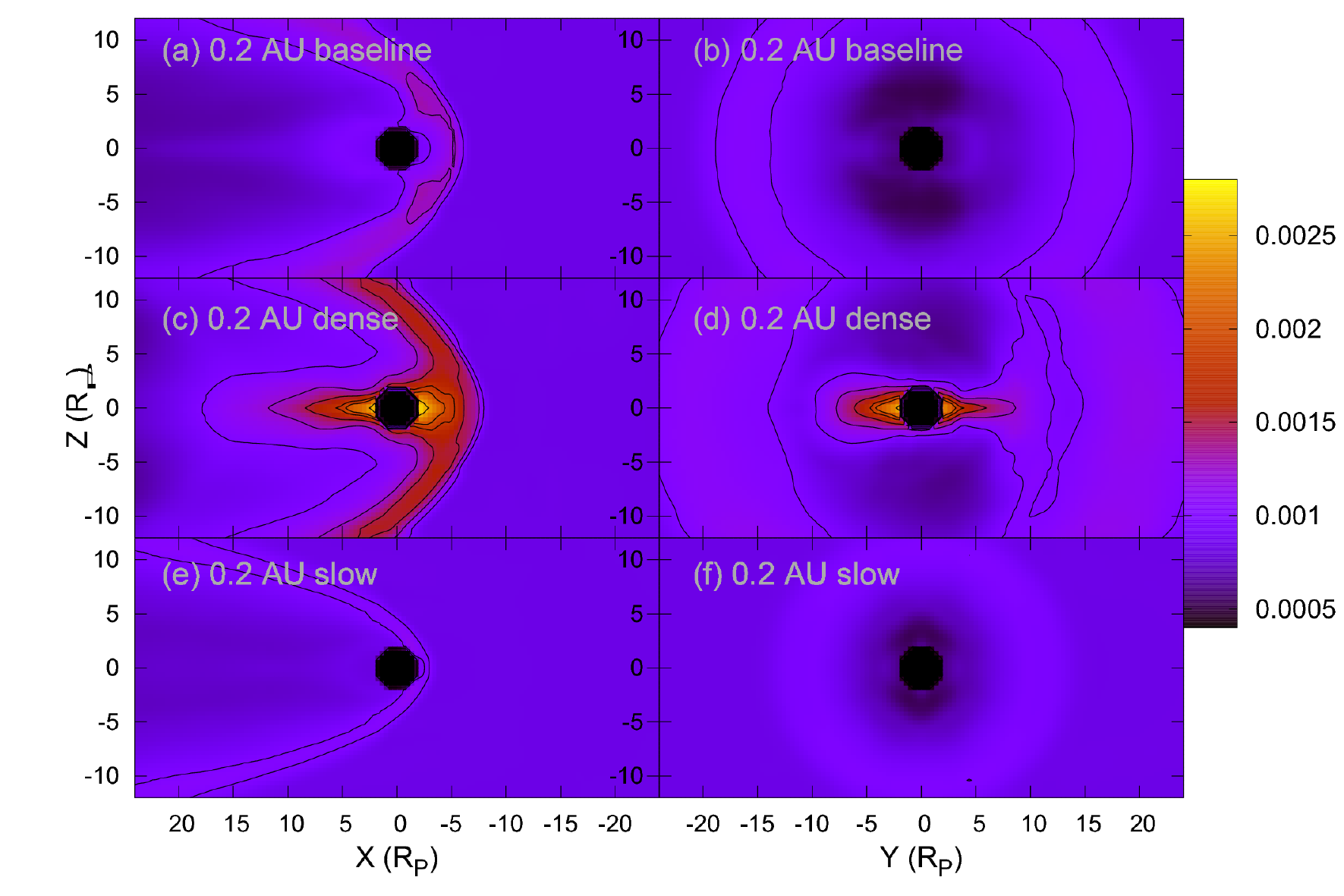}
    \caption{Optical depth, $\tau$, for stellar emitted Mg-II predicted for (a) and (b) the 0.2 AU baseline, (c) and (d) the 0.2 AU dense case, and (e) and (f) the 0.2 AU slow case. The left column corresponds to the ahead-shock case (likely for hotter orbits), and the right column corresponds to the sub-solar shock case (likely for warm orbits). Contours correspond to the hash marks on the colorbar - $\tau$ = 0.0005, 0.001, 0.0015, 0.002, and 0.0025.}
	\label{fig:opticaldepth}
\end{figure*}

One can run a simple calculation to estimate the power reaching Earth for such a distant source. We assume isotropic emission, a 400 kHz frequency, and use the idealized power calculations from above for a Saturn-like planet in a 0.2 AU orbit, at a distance of 10 parsecs from Earth. The signals reaching us would be $\sim$0.19 mJy and 0.29 $\mu$Jy for the best-case, kinetically-controlled 7.2$\times$10$^{10}$W emission and worse-case, magnetically driven 1.1$\times$10$^8$W emission, respectively. The former is at the threshold of current radio instrumentation, and the latter is beyond the technological horizon. This calculation explicitly ignores sources of noise and dispersive effects due to interstellar material.  

For such a close semi-major axis, it is unlikely that direct observation (optical, IR, UV) could occur for said planet. Transit and radial velocity methods are strong at that distance, given a very narrow window of orbital inclinations - if there exists even a moderate amount of orbital inclination relative to our line of sight, then these methods will not be reliable. However, if the combination of orbital and rotational inclinations was such that our line of sight was anywhere in the southern hemisphere of the planet, it is likely that some of the planetary emitted radio power could be detectable with near constant visibility and emission (modulated by stellar activity and magnetospheric response). Based solely upon availability of emitted radio geometry in the example of Fig.~\ref{fig:exosat_radio}, the potential number of planets observable relative to those constrained by orbital configurations amenable to transit or radial velocity measurement could dramatically increase with the use of radio observation. Granted, radio observation has its own set of difficulties to overcome (e.g. need for a large interferometric baseline, Earth's ionosphere). 

The simulated data in Fig.~\ref{fig:exosat_radio} is highly simplified, as we don't simulate the kinetic processes involved in the radio emissions. Our model does not have the ability to predict absolute spectral flux density or more exact radio auroral locations and geometry. The figure merely suggests that the stellar wind dynamic ram pressure and magnetospheric dynamics can directly control not only the field aligned current systems related to power output by ARE, but likewise the celestial coverage of such emissions will be altered for stronger forcing. In the continuum of planetary configurations, some similarity is maintained, while differences arise from the increased dynamic ram pressure; this simple analysis offers an initial view at predicting and characterizing the future observations of radio emissions by giant exoplanets once basic facts about the magnetic properties are observed.

\subsection{Implications for UV transit modification}
\label{sec:optdepth}
There is potential for the increased plasma densities in both the satellite-generated plasma tori and the bow shock of a warm or hot planets to affect transit light curve observations. \citet{benbal14} showed that the inclusion of satellite-generated plasma tori could contribute to the signatures of early ingress for transit observations for both HD 189733 b and WASP-12b. Observations have also been made with the Hubble Space Telescope (HST) that suggest increased bow shock density for these planets has potentially modified transit signals, but other effects - such as absorption by a dense Roche lobe or systematic uncertainties - have not been ruled out \citep[e.g.][]{llama11,llama13,nichols15,alexander16}. 

Unlike the work above on WASP-12b and HD 189733 b, we have assumed our warm-Saturn is orbiting a Sun-like star, though at a cooler orbit of 0.2 AU. This situation is similar to the exoplanet HD 33283 b (as well as numerous unconfirmed Kepler objects of interest (KOI)), orbiting its G3V host with a semi-major axis of 0.168 AU \citep{johnson06}, though transit is not the method of discovery for this planet. Given a similar stellar type, we assume the composition of the host of our star is like that of the Sun; in particular, the ratio of Mg to H,

\begin{gather}
    \frac{n_{Mg}}{n_H} \simeq 6.76 \times 10^{-5}, \label{eq:mgfrac}
\end{gather}

\noindent is taken to be the same as the reported ratio in solar abundance \citep{grevesse07}. 

Using the above assumptions, we have calculated the optical depth for the \ion{Mg}{2} doublet at 279.55 and 280.27 nm in the magnetosphere of our hypothetical warm Saturn, given by 

\begin{gather}
    \tau = 4 \int n_{Mg \textsc{ii}}\, \sigma_{Mg \textsc{ii}}\, dS,
\end{gather}

\noindent where $n_{Mg \textsc{ii}}$ is the number density of \ion{Mg}{2} ion relative to the number density of protons in the simulation given by Eq.~\ref{eq:mgfrac}, $\sigma_{Mg \textsc{ii}}$ is the extinction cross-section of Mg \textsc{ii}, taken $\sigma_{Mg \textsc{ii}} = 6.5 \times 10^{-14}$ cm$^2$, as given in \citet{llama11}, with detailed background in \citet{lai10}. It should be noted that the opacities calculated using this method for the magnetosheath region are reliant upon the plasma temperature remaining at $\sim$10$^4$ K ($\sim$0.86 eV) which is highly unlikely in such systems. Temperatures in the present work, for instance, reach up to $\sim$5 keV ($\sim$5.8$\times$10$^7$ K)in the magnetosheath at 0.2 AU. However, the plasma torus temperatures are a few orders of magnitude lower, at $\sim$30 eV.  

The optical depth absorption profiles for an ahead-shock are shown in the left column of Fig.~\ref{fig:opticaldepth}. The top panel shows the base case for our warm exo-Saturn at 0.2 AU, in which the highest values of optical depth reach $\sim$0.0013, and is located in the interior of the magnetosphere and just outside in the magnetosheath, generated by the pileup of the shocked stellar wind. The plasma torus contributes weakly. The middle panel shows the 0.2 AU dense case, in which the increased ionization from the stellar host is taken into account as the base case holds all quantities as they are expected at 10 AU. In this case, we see that the increased ionization rates in the upper ionosphere of the planet, as well as the satellite plasma torus contribute more strongly to the optical depth enhancement than the simple pileup of density at the bow shock of the magnetosphere - and the overall optical depth reaches a value of $\sim$0.0027. The heavy ions lost from the ionosphere (in this case assumed to be \ion{Mg}{2}) are contained in the inner magnetosphere, unable to radially propagate as seen in Sec.~\ref{sec:interchange}; the enhanced density from the satellite torus plasma in areas both interior to the magnetopause, and in the magnetosheath and bow shock regions contributes directly to an increase in optical depth in those regions. The orbit of the satellite at 4 R$_P$ ensures the ion producing exomoon is contained by the magnetopause (except in the 0.2 AU slow case), but plasma collects in the boundary region and piles up internally against the magnetopause, creating a strong ion density that flows to high latitudes within the magnetosphere. 

One could assume a plasma torus with a much higher concentration of \ion{Mg}{2}, given volcanic activity, for instance. Singly ionized Mg - at a mass to charge ratio of 24 amu C$^{-1}$ - would behave similarly to the 18 amu C$^{-1}$ injected in the present study. If we assume volcanic input of Mg instead of the water-group ions in our plasma torus, the optical depths calculated in Fig.~\ref{fig:opticaldepth} with a pure \ion{Mg}{2} plasma torus would be quite optically thick ($\tau$ $>$ 1) - particularly for systems like the 0.2 AU dense case, where the torus densities reach a few times $\sim$10$^3$. However, the photon count at the desired wavelengths for \ion{Mg}{2} absorption from a Sun-like star is low, and the presence of other species would be beneficial.  

The potential effect on transit light curve observations from both the plasma torus and the enhanced magnetosheath region is shown in such a configuration. Near the inner boundary of our simulation, at 2 R$_P$, the optical depth reaches values of more than double the highest in the base case, at a level $\sim$0.0027. The shocked region shows an enhancement of $\sim$40\%, with optical depths up to $\sim$0.0018. While these depths are not near the 1\% or so anomalies observed by HST for the hot Jovians \citep{nichols15}, future missions will likely reach lower sensitivity; taking into account a dense plasma torus as discussed above for the 0.2 AU dense case, and that a torus-producing moon will have less restricted conditions on its stability at 0.2 AU than 0.02 AU, the potential for transit modifications from giant magnetospheres in a warm orbit should be considered.

This spreading of the satellite generated heavy ion plasma from the torus region to cover such a large area interior to the magnetopause could indicate a strong signature in the transit light curves, depending upon through which portion of the magnetosphere the stellar flux is observed. The strength of the transit signal modification predicted here is detectable by current technology, and future instrumentation. In this analysis, it is noted that we calculated the optical depth for a doublet line emission for only a single species, and assume an actual observation would encompass many wavelengths across stellar ion species to increase the photon count across the stellar spectrum.

\section{Summary and Conclusion}
\label{sec:summary}

We have simulated the effects of stellar wind dynamic ram pressure on a Saturn-like exoplanet with an Enceladus-like plasma torus using a 3D multifluid plasma model. The simulation at 10 AU sets a baseline, with inputs from Cassini data. All planetary parameters are kept constant, varying only the stellar wind dynamic ram pressure in each case, aside from the two special cases of 0.2 AU dense and 0.2 AU slow. The simulation output suggests the following:

\begin{enumerate}
    \item The planetary magnetosphere for a Saturn-like planet with an Enceladus-like satellite plasma torus becomes stable to centrifugal interchange at some point moving starward from a semi-major axis of 5 AU to 1 AU, given constant ionospheric and satellite torus density. The stability criteria shown in Fig.~\ref{fig:interchange_conditions} support this conclusion. For a Saturn-like system, the ratio of thermal energy to corotational energy crosses the threshold of stability at unity between 5 and 1 AU. 
    \item The present work suggests there is an direct relationship between magnetospheric compression and time scale to reach equilibrium mass loss. Highly compressed magnetospheres (e.g. 0.2 AU and 1 AU in the present work) reach equilibrium mass loss rate rapidly, and are stable to interchange. While the equilibrium value for mass loss is equivalent for each semi-major axis studied, less compressed magnetospheres (e.g. 5 AU and 10 AU) start with a lower outflow rate after quiescent stellar wind passage, but fall within the window of instability for centrifugal interchange. The multifluid simulations suggest that while interchange does not affect the overall mass loss rate of a planetary system, stronger instability (like that at 10 AU, when compared to 5 AU) can reduce the time scale to reach mass outflow equilibrium.
    \item For warmer orbits, the magnitude of auroral-related field aligned-currents (FAC) are increased, with a corresponding broadening of latitudinal spread and equatorward locations. This has implications for planetary auroral radio signatures.
    \item From the perspective of characterizing potential radio observations of exoplanets, the coverage of radio emissions varies broadly with increasing stellar wind dynamic ram pressure and beaming angle (emission frequency). Emissions with wide beaming angles are more impacted from the increase in steady state stellar wind pressure, leading to a large portion of the celestial sphere around the planet having higher radio coverage.
    \item UV transit observations could be impacted by the bow shock and/or satellite generated plasma torus for a  planet at a warmer orbit of $\sim$0.2 AU. This leads to potential planetary characterization for a larger population of giant planets than just hot Jovians.
\end{enumerate}

In the present work, we simulated hypothetical magnetospheres for giant, rapidly-rotating planets in increasingly warmer orbits. All orbits were outside the critical Alfv\'{e}n radius for the Sun-like star, and so no direct star-planet interaction (SPI) was simulated. Future work will include more simulations in the transition region between 5 AU and 1 AU for the Saturn-like planet, a case at 0.2 AU which includes the effects of the dense and slow cases, simultaneously. It would also be interesting to simulate a terrestrial planet analog to compare with the Saturn-like planet, effects of CME events, a more varied IMF configuration, and seasonal effects due to the rotational inclination of the planet. The inclusion of a high-resolution grid around a potentially habitable exomoon orbiting at distances of various planetary radii is also a future goal.

\section{Acknowledgements}

The authors would like to thank the anonymous reviewer for their careful reading and helpful suggestions which improved the quality of this paper. We thank R. Barnes for helpful comments and the use of EQTide code to determine the potential locking state of the Saturn-like planet at 0.2 AU. We also thank V. Meadows and E. Agol for helpful discussions. M.T. was partially supported by NASA grant NNX12AK02G while working on this project. This research has made use of the Exoplanet Orbit Database and the Exoplanet Data Explorer at exoplanets.org.

\begin{figure}
    \centering
    \includegraphics[width=0.8\linewidth]{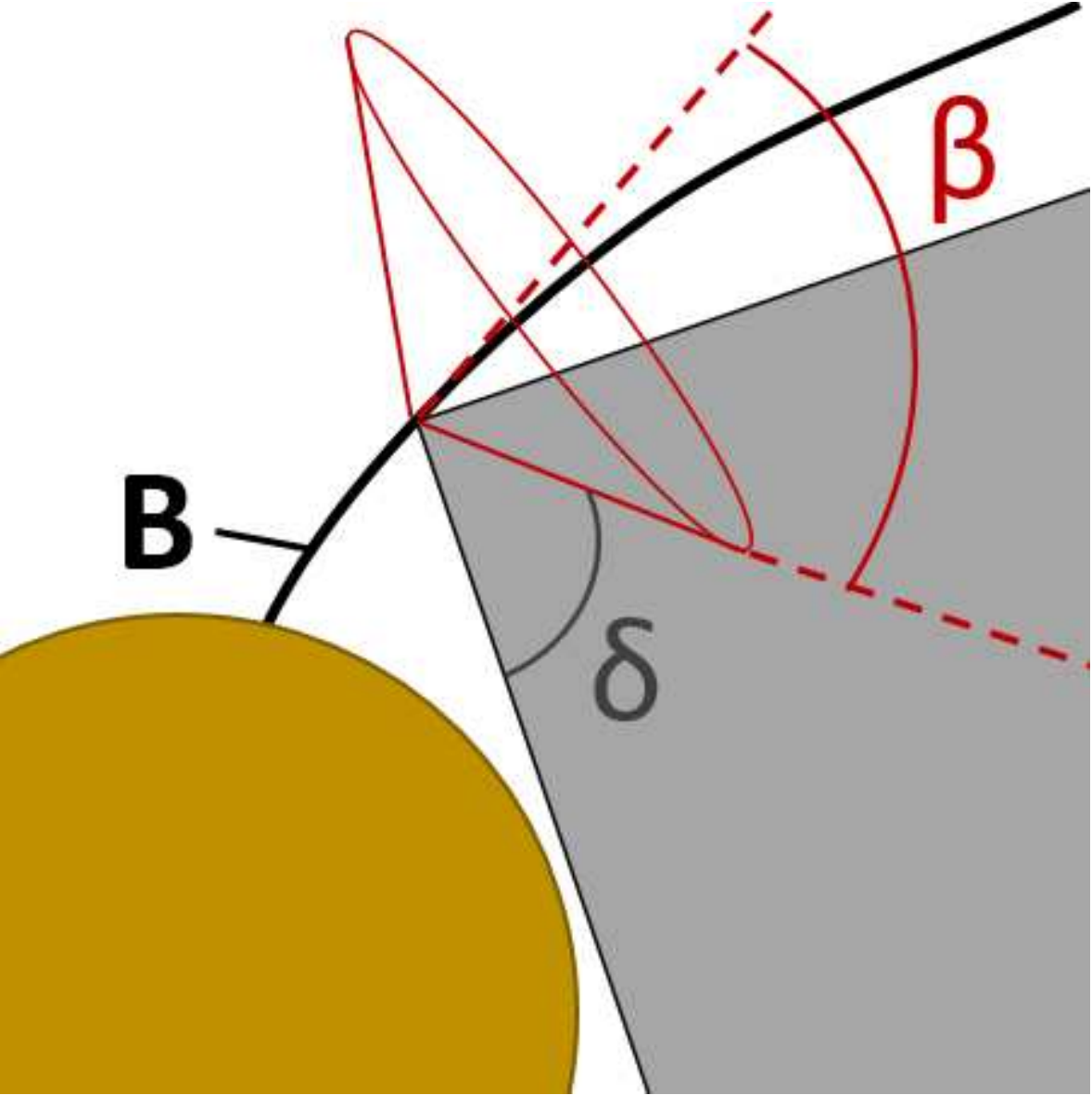}
    \caption{Visualization of radio emission geometry generated along magnetic field line, B, with beaming angle, $\beta$, and soft beaming angle, $\delta$.}
	\label{fig:exosat_geometry}
\end{figure}

\appendix
\section{Radio projection calculation}
\label{app:exosat}

A zeroth order model was developed to estimate the projection of auroral radio emissions on a celestial sphere surrounding the Saturn-like planet as shown in Fig.~\ref{fig:exosat_radio}. The model was developed by using the field-aligned current (FAC) latitudes given by the multifluid model as in Fig.~\ref{fig:polarFAC} and Cassini observations of Saturnian Kilometric Radiation (SKR) as reported by \citet{cecconi09} - including beaming angle and longitudinal measurements. 

A common visualization of radio emission is seen in Fig.~\ref{fig:exosat_geometry}. The red cone represents the typical geometry of emission that has been measured by satellite instrumentation at Saturn. B labels a dipole magnetic field line extending from the planet in the lower left. The beaming angle, or aperture angle, of emission is labeled by $\beta$ and defined as the angle of emission relative to the magnetic field at the point of origin - illustrated by the dashed red lines. $\delta$ denotes the 'soft' beaming angle, shown by the shaded area, that corresponds to power scaling relative to the particle trajectory, or 'on-cone' emission, for particle generated electromagnetic emissions \citep[e.g.][]{rybicki08}.  

The following assumptions were made in our model:
\begin{enumerate}
    \item The surface of the projected sphere plotted in Fig.~\ref{fig:exosat_radio} was located at a distance $\gg$ $\mathcal{O}$({R$_P$}).
    \item Planetary shadowing (projected emission intersecting with and blocked by the planet) was ignored.
    \item Two values were chosen for $\beta$ in both northern and southern hemispheres according to the extreme values given by Cassini observations \citep{cecconi09}. As shown in Fig.~\ref{fig:exosat_radio}, the left column corresponds to $\beta$=90$^{o}$(60$^{o}$) in the northern(southern) hemisphere, and the right column corresponds to $\beta$=65$^{o}$(45$^{o}$) in the northern(southern) hemisphere.
    \item Absolute emitted spectral flux densities (e.g. in units of Jy) are ignored, as this is simply a visualization of radio power projected around the planet as a function of magnetospheric morphology and dynamics. 'On-cone' emission power (along the red dashed lines in Fig.~\ref{fig:exosat_geometry}) was assigned a normalized value of 1, scaling by sin$^2$($\theta$) out to an angle of $\delta$. Angles larger than $\delta$ were not included.
    \item $\delta$ was set to 45$^{o}$ which corresponds to a value of 1/2 maximum power emitted 'on-cone', as power emitted is given by P $\propto$ sin$^2$($\theta$), where $\theta$ is the angle between 'on-cone' emission and point of observation.
    \item Emissions were modeled as being generated between the latitude endpoints given in Fig.~\ref{fig:polarFAC} for the 10 AU and 0.2 AU cases, and between longitudes correlated with Saturn Local Times (SLT) 04:00 and 16:00 \citep{cecconi09} - essentially a 180$^{o}$ coverage in longitude.
\end{enumerate}

Emission cones were modeled on a planetary long-lat grid corresponding to the longitude and latitude boundaries given above, with a spacing of 1$^o$ for latitude, and 1.5$^o$ for longitude. The celestial sphere consisted of a grid in celestial long-lat pairs, with spacing of $\sim$15.7$^o$ for celestial latitude, and $\sim$31.4$^o$ for celestial longitude. For each point in the celestial long-lat projection, the contribution of all emission cones from each planetary long-lat pair were summed and averaged over the total number of cones modeled. This process resulted in each point on the celestial long-lat map representing the projected radio emission relative to the maximum potential of 1.0 - if all emission cones were oriented identically, giving a type of 'heat map' for planetary radio emission.

%\begin{figure}
%    \centering
%    \includegraphics[width=\linewidth]{plot_fac}
%    \caption{Simulated time-averaged, upward-flowing, field aligned currents (FAC) in the northern hemisphere at an altitude of 2.2 R$_P$. Peak FACs occur at latitudes 70$^\text{o}$, 67.0$^\text{o}$, 61.0$^\text{o}$, and 56.5$^\text{o}$ for 10, 5, 1, and 0.2 AU, respectively.}
%	\label{fig:FAC}
%\end{figure}
%
%\clearpage

\end{document}